\documentclass[11pt]{article}
\usepackage{amssymb,amsmath,amsthm,latexsym,natbib,amsbsy}
\usepackage{graphicx}
\usepackage{float}
\usepackage{caption}
\usepackage{subcaption}
\usepackage{epsfig}
\usepackage{epstopdf}  
\usepackage{color}
\usepackage[font= {small,it} ]{caption}
\usepackage{url}
\usepackage[titletoc]{appendix}
\usepackage{xcolor}  

\usepackage[margin= 1in]{geometry}

\title{Nonparametric Functional Calibration of Computer Models}
\author{D. Andrew Brown\thanks{Corresponding author, Department of Mathematical Sciences, Clemson University, Clemson, SC 29634, Email: ab7@clemson.edu} \and  Sez Atamturktur\thanks{Glenn Department of Civil Engineering, Clemson University, Clemson, SC 29634}}

\begin{document}
\maketitle

\thispagestyle{empty}

\begin{abstract}
\noindent Standard methods in computer model calibration treat the calibration parameters as constant throughout the domain of control inputs. In many applications, systematic variation may cause the best values for the calibration parameters to change between different settings. When not accounted for in the code, this variation can make the computer model inadequate. In this article, we propose a framework for modeling the calibration parameters as functions of the control inputs to account for a computer model's incomplete system representation in this regard while simultaneously allowing for possible constraints imposed by prior expert opinion. We demonstrate how inappropriate modeling assumptions can mislead a researcher into thinking a calibrated model is in need of an empirical discrepancy term when it is only needed to allow for a functional dependence of the calibration parameters on the inputs. We apply our approach to plastic deformation of a visco-plastic self-consistent material in which the critical resolved shear stress is known to vary with temperature.\\

\begin{keywords}
Bayesian statistics, Gaussian process, identifiability, model validation, uncertainty quantification, visco-plastic self-consistent material
\end{keywords}
\end{abstract}

\newpage

\thispagestyle{empty}

\section{Introduction}

Many physical phenomena studied in engineering and science disciplines are driven by extremely complex processes that may only be partially understood. Experiments are needed to better understand these processes, but conducting them may be difficult due to economic, technical, or ethical limitations. In response to the need to study such prohibitively resource-intensive systems, the use of computer simulations as proxies for physical observations is now common practice. The design and analysis of computer experiments has become a critical tool in the advancement of numerous fields including national defense, environmental protection, medicine, and manufacturing.

The utility of any computer model is contingent upon that model's fidelity to physical reality. Determining whether or not a specific computer code is an acceptable surrogate for reality falls under the purview of model validation and the closely related area of model calibration. The aim of computer model calibration is to find appropriate values of the parameters governing the computer code under which the code will most closely approximate physical observations according to a predefined metric. Standard methods in computer model calibration treat the calibration parameters as fixed (or averaged) values that are constant throughout the domain of control inputs \cite[e.g.,][]{KennedyOhagan01, WilliamsEtAl06, BayarriEtAl07, HigdonEtAl08}. Computer output and physical data then are combined to obtain the posterior distribution of the calibration parameters. The posterior distribution serves as the basis for calibrating the computer code in which the calibration parameters are set to a point estimate such as the posterior mode \citep{KennedyOhagan01} or varied over the plausible range for making predictions of future responses \citep[e.g.,][]{ReeseEtAl04, KennedyEtAl06, HigdonEtAl08}.

Often in practice the best settings for the calibration parameters may change with different settings of the control inputs \citep{FugateEtAl06, AtamEtAl14, PourhabibEtAl15, PlumleeEtAl15}. This variation can be caused by differences between manufacturing runs, raw materials, etc., or systematic variation not accounted for in the computer code due to incomplete knowledge of the physical system or computational difficulties. The former case was considered by \cite{XiongEtAl09}, who used a hierarchical model to treat the calibration parameters as realizations from a common distribution with parameters estimated via maximum likelihood. The purpose of this article is to treat the latter case by modeling the calibration parameters as functions of the control inputs. To fully account for the uncertainty associated with the unknown functional form, we use a Gaussian process prior while allowing for constraints imposed by opinions of subject matter experts, as is conventionally done in computer experiments. Functional calibration is a topic of interest to many science and engineering fields. Recently, \citet{PlumleeEtAl15} presented a case study in which the calibration parameters are similarly modeled with Gaussian process priors to capture functional dependence for the specific application of the ion channel models of cardiac cells. With this paper, we contribute to solving the problem in applications where available experimental data are scarce, in which case the use of expert-elicited prior constraints becomes necessary to address identifiability issues.

Our aim here is to propose a general framework for nonparametrically modeling calibration parameters as smooth functions of the control inputs. We provide guidance for implementing our so-called state-aware calibration by discussing practical computational considerations, identifiability issues, and determining when to invoke state-aware analysis. We demonstrate the feasibility and performance of our model through an extensive simulation study as well as an application to plastic deformation of a visco-plastic self-consistent (VPSC) material in which the critical resolved shear stress varies with temperature.

The remainder of this paper is organized as follows: In Section 2 we briefly review existing approaches, including the framework of \cite{KennedyOhagan01}, and state our proposed nonparametric functional calibration model. We explicate a special case of our model relevant to the VPSC application, including a discussion of computational considerations when implementing the model via Markov chain Monte Carlo. This is followed by a simulation study in Section 3 comparing our model under different sets of prior constraints with a model assuming a known parametric functional form of the dependence, and with a model that treats all calibration parameters as fixed throughout the experimental domain. We apply our proposed model to the VPSC problem in Section 4. We conclude with discussion of these results, suggestions for determining when functional calibration is necessary, and thoughts about future research in Section 5.

\section{Methods}\label{sec:Methods}
\subsection{General Formulation}
A key reference for the following development is the seminal work of \cite{KennedyOhagan01}, but the notions of model validation and calibration appear at least as early as \citet{BermanNagy83} and \cite{Park91}. Early Bayesian perspectives on calibration may be found in \cite{CraigEtAl01} and \cite{ReeseEtAl04}, with a maximum likelihood approach being presented in \citet{LoeppkyEtAl06}. Methods for integrating field data and computer output for calibration and analysis were presented in \cite{HigdonEtAl04} and \cite{WilliamsEtAl06}. \cite{BayarriEtAl07} suggested a general framework for the model validation process, including calibration. Computer models with high-dimensional output were calibrated using basis function representations of the output in \cite{HigdonEtAl08}. \cite{JosephMelkote09} modified the approach of \cite{KennedyOhagan01} to separate estimation of calibration parameters from determination of a functional form for the model discrepancy. The determination of appropriate values of tuning parameters and calibration parameters simultaneously was done in \cite{HanEtAl09}. Calibration methodology was extended to computer models for nonstationary spatiotemporal processes in \cite{PratolaEtAl13}. \citet{TuoWu2015} discussed calibration based on $L_2$ projections and studied the estimators' asymptotic properties compared to the method of ordinary least squares. \citet{PourhabibEtAl15} treated the calibration parameters as latent variables and used monotone sums of splines to represent the functional relationship between the latent variables and control inputs. Nonparametric functional calibration ideas appear also in \citet{AtamBrown15} and \citet{PlumleeEtAl15}.

Suppose we have $N$ field observations taken at experimental design settings $\mathbf{x}_1, \ldots, \mathbf{x}_N$, where $\mathbf{x}_i \in [0, 1]^{d_x}, ~i= 1,\ldots, N, ~d_x \geq 1$. Denote the field data as $y_i = y(\mathbf{x}_i), ~i= 1, \ldots, N$. Let $\eta(\mathbf{x}, \mathbf{t})$ denote the output of the approximating computer code using control input $\mathbf{x}$ and calibration parameter input $\mathbf{t}$. Here we assume that the computer code is fast-running so that a surrogate is not needed to emulate the computer output. Suppose that any discrepancy between the computer output and the field data is solely due to misspecified parameters in the computer model and measurement error. The field data then can be modeled as
\begin{equation}
y_i = \eta(\mathbf{x}_i,\boldsymbol{\theta}) + \varepsilon_i, ~~i= 1, \ldots, N,\label{eqn:basicModel}
\end{equation}
where $\boldsymbol{\theta}$ is the vector of true parameter values under which the computer model agrees with reality. We assume that $\boldsymbol{\varepsilon} = (\varepsilon_1, \ldots, \varepsilon_N)^T \sim N_N(\mathbf{0}, ~\lambda_{y}^{-1}\mathbf{I})$, where $\mathbf{I}$ is the identity matrix and $\lambda_y > 0$.

Now consider a situation in which $\boldsymbol{\theta}$ depends on the particular settings of the experiment \citep[e.g.,][]{XiongEtAl09, AtamEtAl14, PourhabibEtAl15, PlumleeEtAl15}. In the case $\text{dim}(\boldsymbol{\theta}) > 1$, we partition the calibration parameters as $\boldsymbol{\theta}(\mathbf{x}) = (\boldsymbol{\theta}_1^T(\mathbf{x}), ~\boldsymbol{\theta}_2^T)^T$, where $\boldsymbol{\theta}_1(\cdot) = (\theta_{11}(\cdot), \ldots, \theta_{1p}(\cdot))^T$ is the vector of state-aware calibration parameters. Suppose {\em a priori} that $\boldsymbol{\theta}_1(\cdot)$ is independent of $\boldsymbol{\theta}_2$. To accommodate the dependence of $\boldsymbol{\theta}_1$ on $\mathbf{x}$ without assuming a functional form of the dependence, we use a nonparametric model for the components of $\boldsymbol{\theta}_1(\mathbf{x})$. Specifically, we appeal to Gaussian process (GP) models \citep{OHagan78, Neal98, SantnerEtAl03}. For $\boldsymbol{\theta}_2$, we follow convention and assign the elements independent uniform priors with ranges to be elicited from subject matter experts.

Assuming independence {\em a priori} among all calibration parameters allows us to define the prior distribution on $\boldsymbol{\theta}$ as $\pi(\boldsymbol{\theta(\mathbf{x})}) = \pi_1(\boldsymbol{\theta}_1(\mathbf{x}))\pi_2(\boldsymbol{\theta}_2)$, where we assign Gaussian process priors independently to the elements of $\boldsymbol{\theta}_1(\cdot)$. In the absence of prior knowledge and to limit the number of parameters to be estimated, we use Gaussian processes with constant mean functions, which are usually sufficient for interpolating GP models \citep{Neal98, BayarriEtAl07}. We wish to honor the expert-elicited bounds on plausible values for functional parameters as we would under conventional calibration. Hence, we scale all of the computer code inputs to lie in the unit hypercube and connect the functional calibration parameters to the GP models through a known link function mapping the unit interval to the real line, as done in generalized linear models \citep[GLMs;][]{McNeld89}. It is reasonable to assume that the functional calibration parameters vary smoothly over the control inputs, so the relationships can be well approximated by infinitely differentiable functions. Hence, we use a Gaussian correlation function. We have, for $i= 1, \ldots, p,$
\begin{equation}  
        g(\theta_{1i}(\cdot)) \stackrel{\text{indep.}}{\sim} \mathcal{GP}(\mu_{\theta,i}, ~\lambda_{\theta,i}^{-1}R_i(\cdot, \cdot)); ~~
        R_i(\mathbf{x}, \mathbf{x}^{\prime}) = \exp\left\{-4\gamma_{\theta,i}\sum_{k=1}^{d_x}|x_k - x_k^{\prime}|^2\right\},\label{eqn:GPModel}
\end{equation}
where $g:(0,1) \rightarrow \mathbb{R}$ is one-to-one and differentiable, $d_x = \text{dim}(\mathbf{x})$, $\lambda_{\theta,i}$ are the unknown precisions, and $\gamma_{\theta,i}$ controls the smoothness of the sample paths of $\theta_{1i}(\cdot)$. The mean functions $\mu_{\theta,i}$ are taken to be constant and fixed. For instance, if we take $g$ to be the logit link, then we can center the GPs around $\log(0.5/(1-0.5)) = 0$, and likewise for other link functions. Note that if we know {\em a priori} that $\theta_{1i}(\cdot)$ will be bounded away from 0 and 1 with high probability, then we can take $g(\theta_{1i}) = \theta_{1i}$ as an approximation to the response function. In Section 3, we compare the performance of the logit, $g(z) = \log(z/(1-z))$, probit (inverse Gaussian distribution function), $g(z) = \Phi^{-1}(z)$, cumulative log-log (c-log-log), $g(z) = \log(-\log(z))$, and identity, $g(z) = z$, functions on a simulated example and show that they are all comparable.

When stronger plausible limits are known for the calibration parameters at certain input settings, we can modify the preceding model to
\begin{equation}\label{eqn:cstrdGP}
    g(\theta_{1i}(\cdot)) \stackrel{\text{indep.}}{\sim} \mathcal{GP}(\mu_{\theta,i}, ~\lambda_{\theta,i}^{-1}R_i(\cdot, \cdot))\prod_{c \in C_i} I(L_{c} < \theta_{1i}(\mathbf{x}_{c}) < U_{c}), ~~i=1,\ldots,p,
\end{equation}
for finite sets of constraints indexed by $C_i$ with $L_{c}, U_{c}$ being the appropriate bounds and $I(\cdot)$ the indicator function. To simulate from such distributions in practice, one can draw from the unrestricted sample paths and discard those not satisfying the constraints. We find this approach to be quite feasible and easy to implement for both our simulations and VPSC application.

The model is completed by specifying priors for the hyperparameters in each Gaussian process. For hyperpriors on the parameters governing the covariance structure of the GP, it is convenient to parameterize the correlation function as $\rho_{\theta,i} = e^{-\gamma_{\theta,i}}$ and to assign $\rho_{\theta,i}$ independent Beta priors, $\rho_{\theta,i} \stackrel{\text{iid}}{\sim} \text{Beta}(1, b_{\theta}), ~i= 1, \ldots, p$ \citep{WilliamsEtAl06}. The shape parameter $b_{\theta}$ is chosen to place most probability mass near one to enforce the assumed smoothness {\em a priori}, say $b_{\theta}= 0.1$ or $b_{\theta} = 0.2$. We take $\lambda_{\theta,i} \stackrel{\text{iid}}{\sim} \text{Ga}(a_{\theta}, b_{\theta})$. If we take $g$ to be the identity function in (\ref{eqn:GPModel}), then we can choose $a_{\theta}$ and $b_{\theta}$ to place the prior probability mass around one, since the calibration parameters are scaled. The precision is not as obvious when modeling the GP on the logit, probit, or c-log-log scale, in which case we can take, e.g., $a_{\theta} = 0.01$ and $b_{\theta} = 0.01$ so that the prior is centered at one with standard deviation $\sqrt{0.01/0.01^2} = 10$. Similarly, we take the error precision to be $\lambda_y \sim \text{Ga}(a_y, b_y)$. Since the data are standardized when calibrating the computer code, we again choose the parameters to concentrate the density near one.

A common problem in computer model calibration is that of identifiability of the calibration parameters \citep{BayarriEtAl07}. Bayesian modeling can mitigate this problem through informative prior distributions \citep{GelfandSahu99, Gustafson05}. In our case, the GP induces correlation between $\boldsymbol{\theta}_1(\mathbf{x}_i)$ and $\boldsymbol{\theta}_1(\mathbf{x}_j), ~\mathbf{x}_i \neq \mathbf{x}_j$, so that they are allowed to share information in determining plausible values in the posterior. However, GP models tend to be erratic near the boundaries of the domains over which they are studied, and this behavior can limit the Bayesian learning about $\boldsymbol{\theta}_1(\cdot)$ or $\boldsymbol{\theta}_2$ in the posterior. Another consequence of weak identifiability is the possibility of highly correlated draws in the MCMC sampling routine, potentially leading to very poor convergence properties. A possible solution is to elicit informative prior distributions from subject matter experts. If an informative prior distribution can be elicited for $\boldsymbol{\theta}_2$, or if the possible sample paths of $\boldsymbol{\theta}_1(\cdot)$ can be constrained using prior information, identifiability can be improved. We return to this point in Section 3.

A goal of computer model calibration is to facilitate reliable predictions at untested experimental settings. In the Bayesian paradigm, such predictions are based on the posterior predictive distribution. Suppose we have training data $\mathbf{y} = (y(\mathbf{x}_1), \ldots, y(\mathbf{x}_N))^T$ and we wish to make predictions for future realizations at $m$ untested settings $\mathbf{x}_1^{\ast}, \ldots, \mathbf{x}_m^{\ast}$, $\mathbf{y}^{\ast} = (y^{\ast}(\mathbf{x}_1^{\ast}), \ldots, y^{\ast}(\mathbf{x}_m^{\ast}))^T$. Since $\mathbf{y}^{\ast}$ is determined by $\boldsymbol{\theta}_1^{(\mathbf{x}^{\ast})} := (\boldsymbol{\theta}_1^T(\mathbf{x}_1^{\ast}), \ldots, \boldsymbol{\theta}_1^T(\mathbf{x}_m^{\ast}))^T, ~\boldsymbol{\theta}_2,$ and $\lambda_y$, and {\em a posteriori} information about $\boldsymbol{\theta}_1^{(\mathbf{x}^{\ast})}$ depends on $\mathbf{y}$ only through the posterior distribution of $\boldsymbol{\theta}_1^{(\mathbf{x})} = (\boldsymbol{\theta}_1^T(\mathbf{x}_1), \ldots, \boldsymbol{\theta}_1^T(\mathbf{x}_N))^T$, $\boldsymbol{\rho}_{\theta} = (\rho_{\theta,1}, \ldots, \rho_{\theta,p})^T$, and $\boldsymbol{\lambda}_{\theta}= (\lambda_{\theta,1}, \ldots, \lambda_{\theta,p})^T$, predictions at untested settings are available by drawing $\boldsymbol{\theta}_1^{(\mathbf{x})}, \boldsymbol{\rho}_{\theta}, \boldsymbol{\lambda}_{\theta}$, $\boldsymbol{\theta}_2$, and $\lambda_y$ from the joint posterior distribution, sampling from the distribution of $\boldsymbol{\theta}_1^{(\mathbf{x}^{\ast})} \mid \boldsymbol{\theta}_1^{(\mathbf{x})}, \boldsymbol{\rho}_{\theta}, \boldsymbol{\lambda}_{\theta}$, and then drawing from $\mathbf{y}^{\ast} \mid \boldsymbol{\theta}_1^{(\mathbf{x}^{\ast})}, \boldsymbol{\theta}_2, \lambda_y$. Here, $\pi(\boldsymbol{\theta}_1^{(\mathbf{x}^{\ast})} \mid \boldsymbol{\theta}_1^{(\mathbf{x})}, \boldsymbol{\rho}_{\theta}, \boldsymbol{\lambda}_{\theta})$ is readily available since $\boldsymbol{\theta}_1^{(\mathbf{x}^{\ast})} , \boldsymbol{\theta}_1^{(\mathbf{x})} \mid \boldsymbol{\rho}_{\theta}, \boldsymbol{\lambda}_{\theta}$ follows a multivariate Gaussian distribution.

When an experimenter has more reliable prior information concerning the functional forms of the dependencies of calibration parameters on the control settings, the nonparametric Gaussian process in (\ref{eqn:GPModel}) can be replaced with a parametric function, $\boldsymbol{\theta}(\mathbf{x}) = f(\mathbf{x}, \boldsymbol{\beta})$. The problem then is to assign an appropriate prior distribution to $\boldsymbol{\beta}$ and estimate plausible values from the posterior distribution. This was done in \cite{XiongEtAl09} and \cite{AtamEtAl14}, with a similar approach taken in \citet{PourhabibEtAl15}. The parametric calibration problem can be expressed as a standard calibration approach, though, since the calibration parameters are still treated as constant and appear in the ``augmented" computer code, $\eta(\mathbf{x}, \boldsymbol{\theta}(\mathbf{x})) = \eta(\mathbf{x}, f(\mathbf{x}, \boldsymbol{\beta})) \equiv \eta(\mathbf{x}, \boldsymbol{\beta})$.\\

\subsection{Two Parameter Model with Scalar Control Input}\label{sub:propModel}
\indent Our motivating example of modeling plastic deformation of viscoplastic self-consistent material involves a single control input and two calibration parameters so that $p=1$ and $d_x = 1$ in (\ref{eqn:GPModel}). In light of this, it is also the scenario we consider in our simulation study in Section 3. We focus on this special case and consider the details more carefully, including specification of the model, the joint posterior distribution, and computational considerations for implementation.

Let $\mathbf{y}= (y(x_1), \ldots, y(x_N))^T$ be the vector of observed field data, $\mathbf{x} = (x_1, \ldots, x_N)^T$ the experimental settings under which the data were collected, and $\boldsymbol{\eta}(\boldsymbol{\theta}^{(\mathbf{x})}) = (\eta(x_1, \boldsymbol{\theta}(x_1)), \ldots, \eta(x_N, \boldsymbol{\theta}(x_N)))^T$ the calibrated computer output at these experimental settings. For ease of notation, we suppress the constraints in (\ref{eqn:cstrdGP}) so that the sample path restrictions are implied. Our proposed model becomes
\begin{equation}
    \begin{aligned}
        \mathbf{y} \mid \boldsymbol{\theta}^{(\mathbf{x})}, ~\lambda_y &\sim N_N(\boldsymbol{\eta}(\boldsymbol{\theta}^{(\mathbf{x})}), ~\lambda_y^{-1}\mathbf{I})\\
        \lambda_y &\sim \text{Ga}(a_y, ~b_y), ~~a_y, b_y > 0\\
        g(\theta_1(\cdot)) \mid \lambda_{\theta}, ~\rho_{\theta} &\sim \mathcal{GP}(\mu_{\theta}, ~\lambda_{\theta}^{-1}R_{\rho_{\theta}}(\cdot, \cdot)), ~~-\infty < \mu_{\theta} < \infty\\
        \theta_2 &\sim \text{Unif}(0, 1)\\
        \lambda_{\theta} &\sim \text{Ga}(a_{\theta}, ~b_{\theta}), ~~a_{\theta}, b_{\theta} > 0\\
        \rho_{\theta} &\sim \text{Beta}(1, ~b_{\theta}), ~~b_{\theta} > 0,
    \end{aligned}\label{eqn:2PModel}
\end{equation}
where $g$ is a known link function as in (\ref{eqn:GPModel}), $\boldsymbol{\theta}_1^{(\mathbf{x})} = (\theta_1(x_1), \ldots, \theta_1(x_N))^T$, and $R_{\rho_{\theta}}(\cdot, \cdot)$ is the correlation function given by $R_{\rho_{\theta}}(x, ~x^{\prime}) = \rho_{\theta}^{4(x - x^{\prime})^2}$. Letting $g(\boldsymbol{\theta}_1^{(\mathbf{x})}) = (g(\theta_1(x_1)), \ldots, g(\theta_1(x_N)))^T$, the joint posterior distribution is then
\begin{eqnarray*}
    \pi(\boldsymbol{\theta}_1^{(\mathbf{x})}, \theta_2, \rho_{\theta}, \lambda_{\theta}, \lambda_y \mid \mathbf{y}) &\propto& \lambda_y^{N/2}\exp\left\{-\frac{\lambda_y}{2}(\mathbf{y} - \boldsymbol{\eta}(\boldsymbol{\theta}_1^{(\mathbf{x})}, \theta_2))^T(\mathbf{y} - \boldsymbol{\eta}(\boldsymbol{\theta}_1^{(\mathbf{x})}, \theta_2))\right\}\lambda_y^{a_y-1}\exp(-b_y\lambda_y)\\
        & &\times \lambda_{\theta}^{N/2}|\mathbf{R}_{\rho_{\theta}}|^{-1/2}\exp\left\{-\frac{\lambda_{\theta}}{2}(g(\boldsymbol{\theta}_1^{(\mathbf{x})}) - \mu_{\theta}\mathbf{1})^T\mathbf{R}_{\rho_{\theta}}^{-1}(g(\boldsymbol{\theta}_1^{(\mathbf{x})}) - \mu_{\theta}\mathbf{1})\right\}\\
        & &\times \lambda_{\theta}^{a_{\theta}-1}\exp(-b_{\theta}\lambda_{\theta})(1-\rho_{\theta})^{b_{\theta}-1},
\end{eqnarray*}
where $\mathbf{R}_{\rho_{\theta}} = \{R_{\rho_{\theta}}(x_i, x_j)\}_{i,j=1}^N$ and $\mathbf{1} = (1, \ldots, 1)^T$.

We use a Markov chain Monte Carlo algorithm \citep[MCMC;][]{GelfandSmith90} to simulate draws from the posterior. To eliminate the boundary constraints on $\theta_2$ and $\rho_{\theta}$, and to make our sampling algorithm less sensitive to the scale of the data, we reparameterize with $\xi = \log(-\log(\theta_2))$ and $\nu = \log(-\log(\rho_{\theta}))$. Note that $\nu$ then is equivalent to the correlation length parameterization suggested by \cite{Neal98} when implementing MCMC for models with GP priors. The subsequent full conditional distributions necessary for the algorithm are given in the Supplementary Material.

We use Gibbs sampling with Metropolis steps \citep{MetropolisEtAl53, GemanGeman84, Tierney94, CarlinLouis09} for the non-standard distributions. In drawing sample paths of $\theta_1(\cdot)$ with Metropolis proposals, we wish to take advantage of the prior smoothness assumptions. Following the suggestion of \cite{Neal98}, we sample $\boldsymbol{\theta}_1^{(\mathbf{x})}$ using a multivariate Gaussian proposal with correlation matrix dependent upon the current value of $\rho_{\theta}$. That is, to sample from the distribution of $\boldsymbol{\theta}_1^{(\mathbf{x})} \mid \xi, \nu, \lambda_{\theta}, \lambda_y, \mathbf{y}$, on the $k^{\text{th}}$ iteration, we find the spectral decomposition of $\mathbf{R}_{\nu} = \mathbf{U}\boldsymbol{\Lambda}\mathbf{U}^T$ and draw a proposal as $\boldsymbol{\theta}_1^{(\mathbf{x}), \dag} = c\mathbf{U}\boldsymbol{\Lambda}^{1/2}\mathbf{z} + \boldsymbol{\theta}_1^{(\mathbf{x}),(k-1)}$, where $\mathbf{z} \sim N_N(\mathbf{0}, \mathbf{I})$ and $c$ is determined adaptively during the burn-in period by monitoring the acceptance rate and adjusting periodically. Note that we use the spectral decomposition of $\mathbf{R}_{\nu}$ despite the fact that it is slower to compute than the usual Cholesky decomposition, since it is more computationally stable for generating Gaussian random variables. For $\xi$, we use a Metropolis step with candidates $\xi^{\dag} \sim N(\xi^{(k-1)}, ~c_{\xi}^2)$, where $c_{\xi}$ is tuned adaptively, and similarly for $\nu$.

When the observed design points are close together, the columns of the correlation matrix $\mathbf{R}_{\nu}$ are nearly linearly dependent so that $\mathbf{R}_{\nu}^{-1}$ is computationally unstable. While the spectral decomposition mitigates the problem when simulating multivariate Gaussian draws, this technique is not helpful in solving the matrix or finding its determinant. To address this, we add a nugget $\delta$ to obtain $\mathbf{R}_{\nu, \delta} := \mathbf{R}_{\nu} + \delta\mathbf{I}$. \cite{RanjanEtAl11} proposed determining the nugget with $\delta = \max\{\lambda_N(\kappa(\mathbf{R}_{\nu}) - e^{a})(\kappa(\mathbf{R}_{\nu}))^{-1}(e^{a} - 1)^{-1}, ~0\}$, where $\lambda_N$ is the largest eigenvalue of $\mathbf{R}_{\nu}$, $\kappa(\mathbf{R}_{\nu})$ is the condition number, and $e^a$ is the threshold on $\kappa(\mathbf{R}_{\nu})$ for the matrix to be well-conditioned. We find that $a = 20$ works well. We use the Cholesky factorization of the modified matrix, $\mathbf{R}_{\nu, \delta} = \mathbf{L}_{\delta}\mathbf{L}_{\delta}^T$, to approximate $\log(|\mathbf{R}_{\nu}|^{-1/2}) \approx -\sum_{i=1}^N \log(l_{ii})$, where $l_{ii}$ is the $i^{\text{th}}$ diagonal element of $\mathbf{L}_{\delta}$.

A useful avenue for future research is the exploration of various computational approaches for simulating the posterior distribution when assuming functional dependence between $\boldsymbol{\theta}$ and $\mathbf{x}$. We obtain acceptable results using the above Metropolis-within-Gibbs sampling scheme, but we still find the smoothness hyperparameter $\rho_{\theta}$ difficult to estimate via posterior inference. Approaches to this problem suggested in the literature include Hamiltonian Monte Carlo \citep{Neal98, Neal11}, or substituting an empirical Bayes estimator such as the posterior mode \citep{QianWu08}. The identifiability problem and the dependence it can induce among calibration parameters in MCMC simulations have motivated useful advances such as adaptive Metropolis sampling \citep[AM;][]{HaarioEtAl01} and delayed rejection adaptive Metropolis \citep[DRAM;][]{HaarioEtAl06}. There is no doubt more research to be done in this area.\\

\section{Simulation Study}\label{sec:Simulation}
To illustrate our proposed method, we simulate field data $y_i, ~i= 1,\ldots, N$, by supposing that $y(x_i) = c_1(x_i) + c_2x_i^2 + \varepsilon_i$, where $\varepsilon_i \stackrel{\text{iid}}{\sim}N(0, ~0.05^2), ~i= 1, \ldots, N$. The computer model is $\eta(x, t_1, t_2) = t_1 + t_2 x^2$ so that both calibration parameters $t_1$ and $t_2$ are assumed constant in the computer code. Suppose that, in reality, $c_2 = 2.5$ is constant across the domain and that $c_1(\cdot)$ is determined by $c_1(x) = 2\sqrt{x}$. The simulated field data are generated at $\mathbf{x} = (0.00, 0.05, 0.10, \ldots, 0.90, 0.95)^T$. To evaluate predictive performance, the responses $\mathbf{y}^{\ast} = (y(0.45), \ldots, y(0.65))^T$ are held out as a validation set, leaving the remaining 15 observations as a training dataset. Note that we are intentionally using a small number of observations, as this is typically the case in practice. We use the logit link, $g(\theta(x)) = \log(\theta(x)/(1-\theta(x)))$, and take $a_{\theta} = b_{\theta} = 0.01$ in the prior on $\lambda_{\theta}$. We set $a_y = b_y = 5$, encouraging $\lambda_y$ to be close to one since the data are standardized. We place most of the prior probability mass near one for $\rho_{\theta}$ with $b_{\theta} = 0.2$. The field data are standardized and the calibration parameters are scaled to lie in the unit hypercube prior to calibrating the computer model; i.e., $\theta_i = (c_i - c_{\text{min},i})(c_{\text{max},i} - c_{\text{min},i})^{-1}, ~i= 1,2$, in (\ref{eqn:2PModel}).

\begin{figure}[tb]
    \centering
    \includegraphics[scale= 0.4, clip= FALSE, trim= 0.25in 0in 0in 0in]{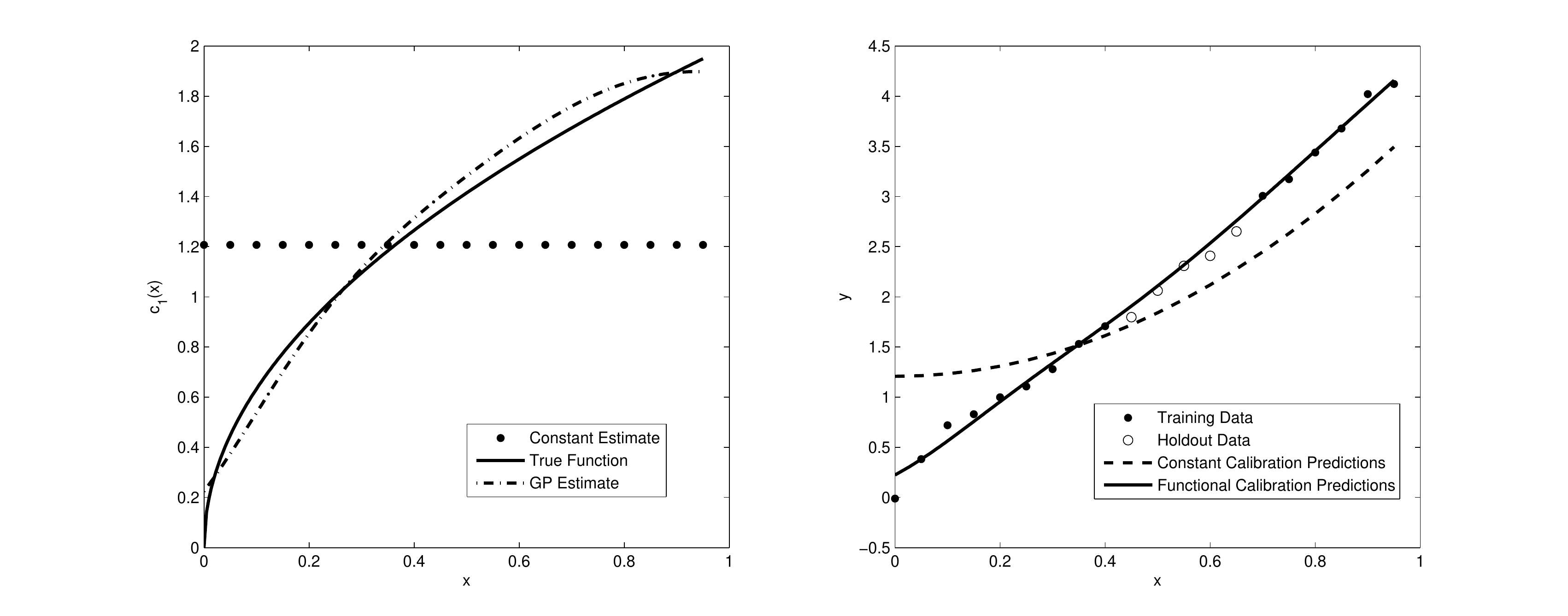}  
    \caption{Simulated data used for calibration under the logit link. All values are plotted on the original scale. In the functional case, we obtain posterior sample paths from $p(\theta_1(\cdot) \mid \mathbf{y}) = \int_{(\boldsymbol{\theta}_1^{(\mathbf{x})}, \lambda_{\theta}, \rho_{\theta})}p(\theta_1(\cdot) \mid \boldsymbol{\theta}_1^{(\mathbf{x})}, \lambda_{\theta}, \rho_{\theta})\pi(\boldsymbol{\theta}_1^{(\mathbf{x})}, \lambda_{\theta}, \rho_{\theta} \mid \mathbf{y})d\boldsymbol{\theta}_1^{(\mathbf{x})}d\lambda_{\theta}d\rho_{\theta}$.  The right panel compares the corresponding predictions versus both the training data and holdout data.}\label{fig:origData}
\end{figure}

To illustrate what is at stake, Figure \ref{fig:origData} plots the simulated data along with the computer model predictions obtained when using the posterior means of the calibration parameters as the estimates inside the code. The left panel plots the estimated posterior means using both a constant assumption on $\theta_1$ (with a uniform prior) as well as the functional assumption with the logit link in which, {\em a priori}, $\log(\theta_1(\cdot)/(1-\theta_1(\cdot))) \sim \mathcal{GP}(0, \lambda_{\theta}^{-1}R_{\rho_{\theta}}(\cdot, \cdot))$.  The circles represent the holdout data. When treating both calibration parameters as constant, the calibrated code yields strong disagreement between the predictions and field data. In practice, this disagreement would likely be absorbed by adding an extra term to (\ref{eqn:basicModel}) to represent model discrepancy. Such an approach would conceal the true nature of the system, illustrated in the left panel of Figure \ref{fig:origData}. By allowing $\theta_1(\cdot)$ to change over the experimental settings, reality is represented in a manner more consistent with experiments without resorting to a purely empirical discrepancy term. Our approach thus allows what would be a previously unknown functional form to emerge. We discuss these results further below. Notably, we show that treating $\theta_1$ as constant still results in posterior concentration about the ``average" value, so that a researcher could gain a false sense of security in their assumptions.\\
\indent We simulate draws from the posterior via Markov chain Monte Carlo using the techniques described in Section 2. For each of the scenarios considered, we run three different chains in parallel using different starting values to assess convergence. Each chain uses a burn-in period of 5,000 iterations. During the burn-in period, the scales of the proposal distributions are adjusted every 100 iterations to attain approximate acceptance rates between 40-50\% and 20-25\% for the univariate and multivariate conditional distributions, respectively \citep{GelmanEtAl95}. After burn-in, the chains are run for an additional 4,000 iterations with the widths of the proposal distributions held fixed. In each chain, every second draw is saved from the sampling loop to reduce autocorrelation. Trace plots are examined to assess convergence, after which the draws for the three chains are combined for a final Monte Carlo sample size of 6,000.

\begin{figure}[tb]
    \centering
    \includegraphics[scale= 0.43, clip= TRUE, trim= 0.5in 0pt 0in 0in]{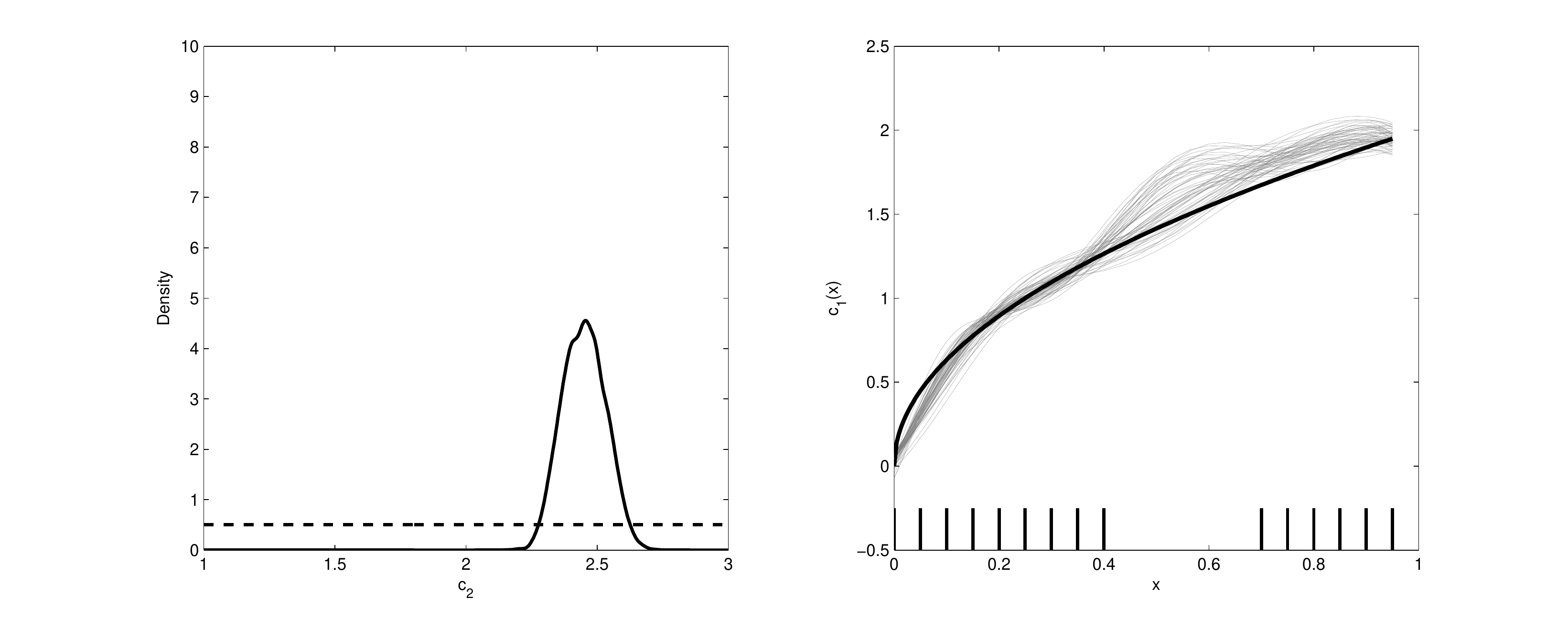}  
    \caption{Posterior draws for $c_1(\cdot)$ and $c_2$ in the two-parameter GP model with constraints on the boundary values of $c_1(\cdot)$ and logit link. The dashed and solid curves in the left panel are the prior and posterior densities of $c_2$, respectively. The thick line in the right panel is the true function $c_1(\cdot)$. The heavy tick marks at the bottom of the right panel indicate the $x$ values of the training data.}\label{fig:strongTauPlots}
\end{figure}

\begin{figure}[tb]
    \centering
    \includegraphics[scale= 0.525, clip= TRUE, trim= 0in 0pt 0in 0in]{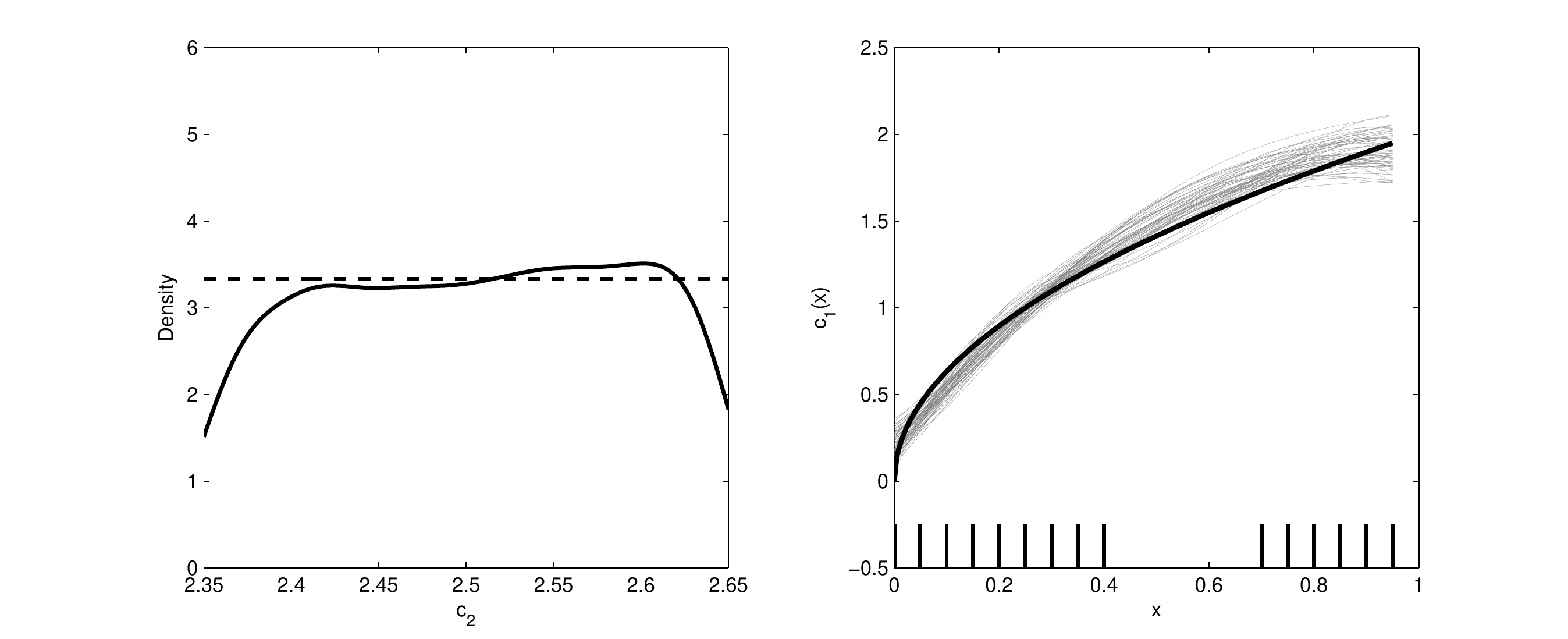}  
    \caption{Posterior draws for $c_1(\cdot)$ and $c_2$ in the two-parameter GP model with logit link and tight prior bounds on $c_2$. The dashed and solid curves in the left panel are the prior and posterior densities of $c_2$, respectively. The thick line in the right panel is the true function $c_1(x)$. The heavy tick marks at the bottom of the right panel indicate the $x$ values used in evaluating the posterior.}\label{fig:strongN0Plots}
\end{figure}

First, we {\em a priori} enforce the constraints $-0.075 \leq c_1(x_1) \leq 0.075$ and $1.85 \leq c_1(x_{20}) \leq 2.05$ so that the constraint on $c_1(x_1)$ has a width of 3 error standard deviations and the constraint on $c_1(x_{20})$ has a width of 4 error standard deviations. By contrast, the prior on $c_2$ is relatively vague. We take it to be uniform between $c_{2, \text{min}} = 1$ and $c_{2, \text{max}} = 3$, so that it measures 40 error standard deviations in width. Figure \ref{fig:strongTauPlots} illustrates the results. In the left panel we see a strong contrast between the prior and posterior densities of $c_2$, demonstrating the considerable Bayesian learning about this parameter that occurred in the posterior. Further, the posterior is correctly concentrating about $c_2 = 2.5$. The right panel plots sample paths from the distribution of $c_1(\cdot) \mid \mathbf{y}$. Superimposed on this plot is the true $c_1(\cdot)$ function. Here we see our model's ability to recover the true functional dependence on $x$, despite $c_1$ being treated as constant inside the computer code. Even though observations are not directly available for individually calibrating $c_1$ at each of the hold out points, the posterior draws tend to closely agree with the truth.\\
\indent The second scenario we consider is the opposite of the first. We place more informative prior bounds on $c_2$ so that it is uniform between $c_{2, \text{min}} = 2.35$ and $c_{2, \text{max}} = 2.65$. We remove the constraints on the values of $c_1(x)$ at any $x$ so that the possible realizations are unrestricted. Figure \ref{fig:strongN0Plots} illustrates the prior and posterior of $c_2$ and sample paths drawn from the posterior distribution of $c_1(\cdot)$. The true functional path of $c_1(\cdot)$ is again plotted for reference. We see the very weak Bayesian learning about $c_2$ that has occurred in this case. This reflects that fact that, given the bounds we have already imposed on the possible values for $c_2$, the data contain little additional information concerning plausible values. We again see posterior concentration of $c_1(\cdot)$ about the true parameter path at both the observed design points as well as at the untested design settings. Thus, in spite of allowing for an unconstrained functional path, we are able to recover the functional form.\\
\indent Suppose we know that $c_1(\cdot)$ can be approximated with $c_1(x) = \beta_0^{U} + \beta_1^{U}\sqrt{x}$, where $\beta_0^{U}$ and $\beta_1^{U}$ are unknown and the superscripts indicate the correspondence with the unscaled parameters. In this case, we alter Model (\ref{eqn:2PModel}) by writing $\boldsymbol{\theta}_1^{(\mathbf{x})} = (\beta_0 + \beta_1\sqrt{x_1}, \ldots, \beta_0 + \beta_1\sqrt{x_N})^T$ and assigning prior distributions to $\beta_0$ and $\beta_1$, where we drop the superscripts to indicate the rescaling. Calibration then involves determining probable values of $\boldsymbol{\beta} = (\beta_0, \beta_1)^T$. To give the data as much freedom as possible in determining appropriate values in this study, we use a flat prior, $\pi(\boldsymbol{\beta}) \propto 1$. Note that this prior is likely to be much weaker than priors used in practice. We use the same prior distribution for $\lambda_y$ as in the previous simulations and again take $c_2$ to be {\em a priori} uniform between $c_{2,\text{min}} = 2.35$ and $c_{2, \text{max}} = 2.65$. We simulate the posterior distribution $\pi(\boldsymbol{\beta}, \theta_2, \lambda_y \mid \mathbf{y})$ via MCMC with the same burn-in period and the same number of chains as with the GP model.

\begin{figure}[h!tb]
    \centering
    \includegraphics[scale= 0.6, trim= 1.25in 0in 0in 0in]{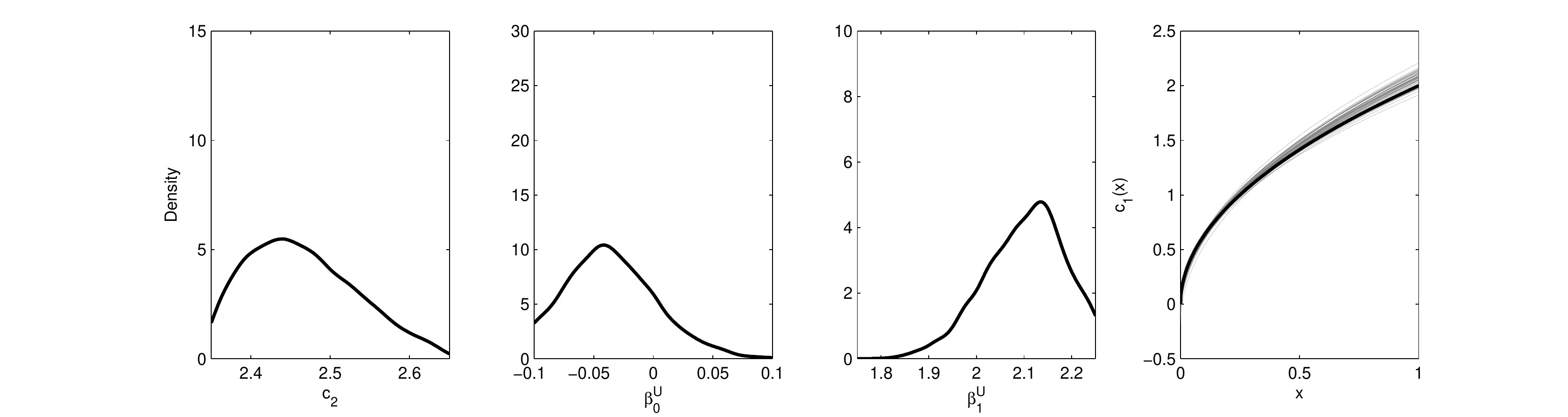}  
    \caption{Smoothed approximate posterior distributions of $c_2$, $\beta_0^{U}$, and $\beta_1^{U}$ (first three panels from the left) when replacing the GP prior with $\theta_1(x) = \beta_0 + \beta_1\sqrt{x}$. The far right panel plots realizations of the estimating curves (grey lines) based on draws of $\beta_0^{U}$ and $\beta_1^{U}$ from their posterior, along with the true function for reference (heavy black line).}\label{fig:fnTauPlots}
\end{figure}

\begin{figure}[h!tb]
    \centering
    \includegraphics[scale= 0.4]{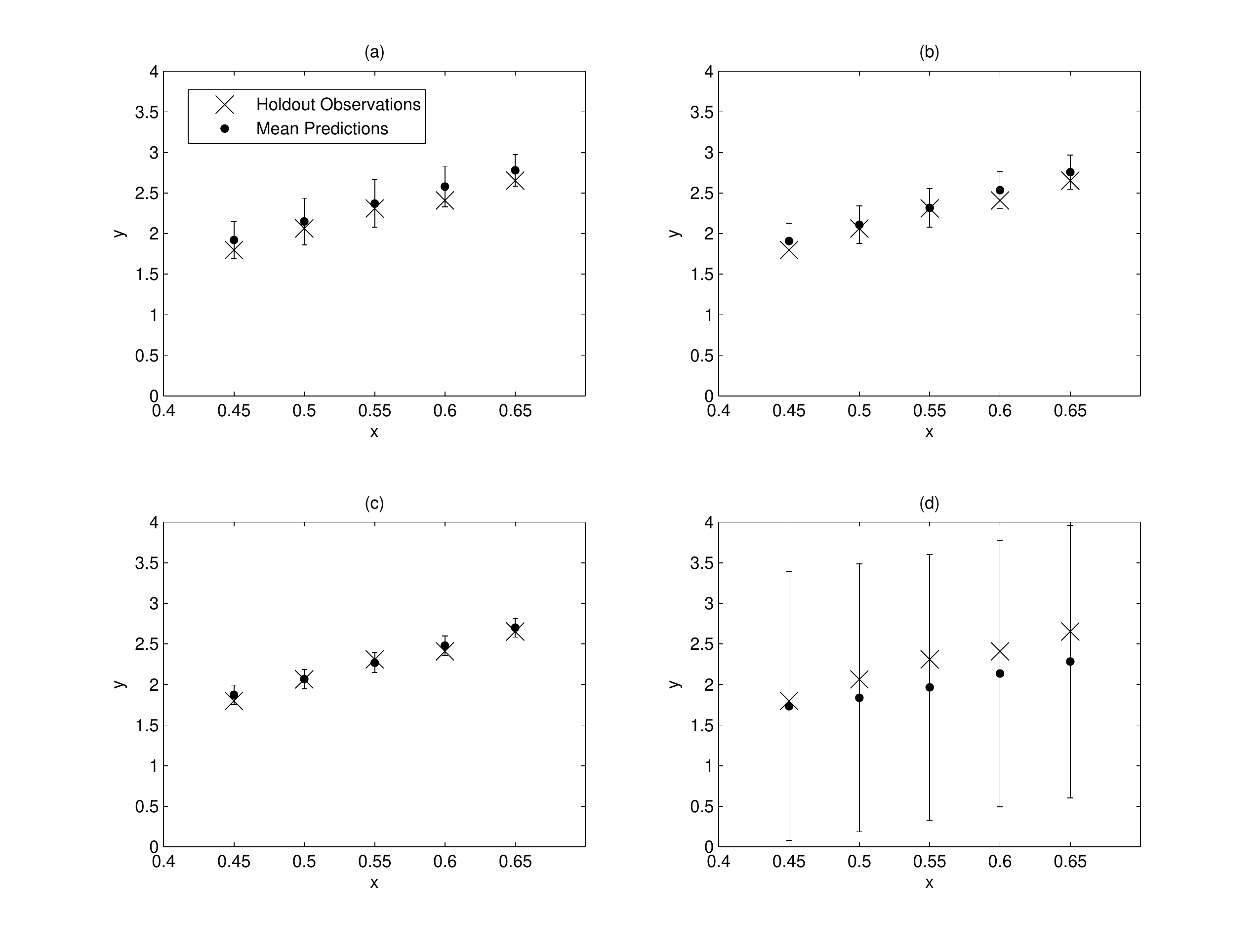}  
    \caption{Posterior predictions at holdout settings with approximate 95\% error bars under (a) $\theta_1(x)$ constrained at $x_1$ and $x_N$, (b) $\theta_2$ constrained between tight prior bounds, (c) $\theta_1(x) = \beta_0 + \beta_1\sqrt{x}$, and (d) $\theta_1$ assumed constant.}\label{fig:4CasesPredPlots}
\end{figure}

\begin{figure}[h!tb]
    \includegraphics[scale= 0.5, clip= TRUE, trim= 0.85in 0pt 0in 0in]{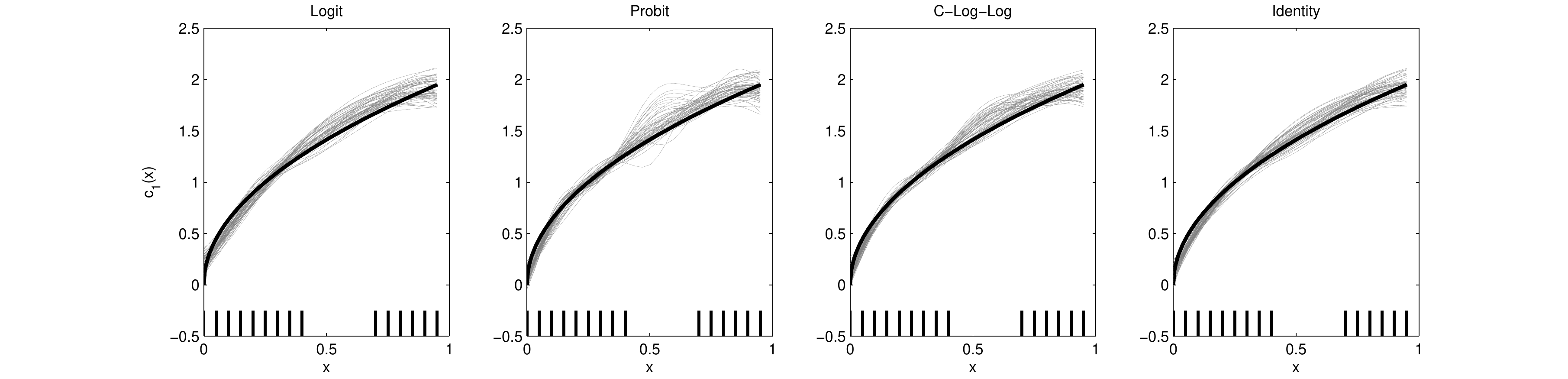}  
    \caption{Posterior sample paths of $c_1(\cdot)$ obtained from using the logit, probit, c-log-log, and identity link functions with the simulated data example.}\label{fig:4LinksPaths}
\end{figure}

\begin{figure}[h!tb]
    \centering
    \includegraphics[scale= 0.5, clip= TRUE, trim= 0.4in 0pt 0in 0in]{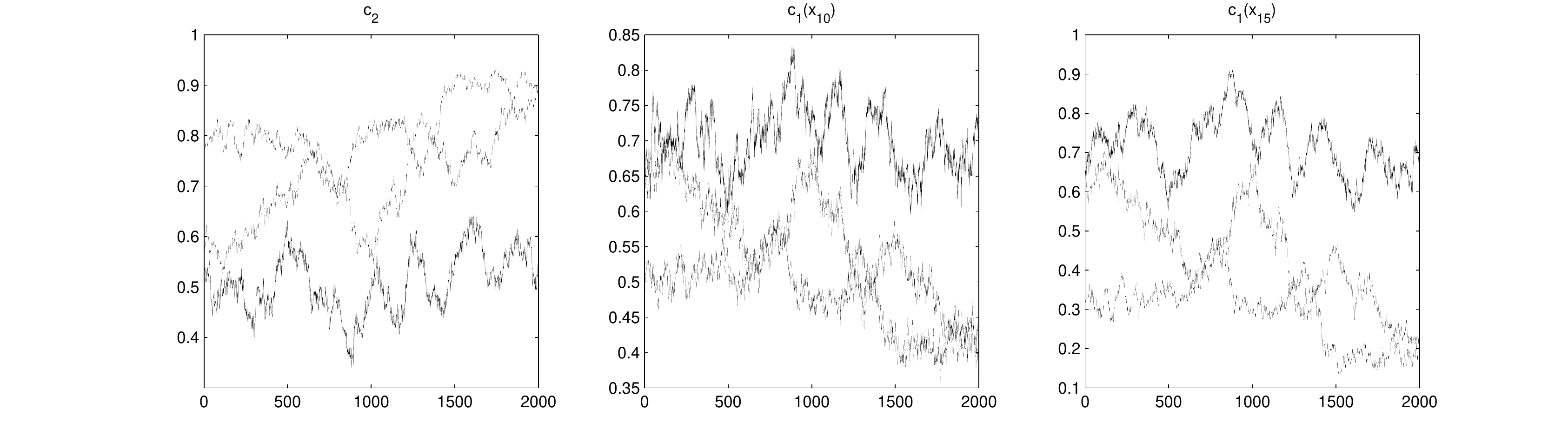}  
    \caption{Trace plots of sampled values of the calibration parameters $c_2$, $c_1(x_{10})$, and $c_1(x_{15})$ for three different chains (with different initial values) under vague priors for both $c_1(\cdot)$ and $c_2$.}\label{fig:BadTraces}
\end{figure}

\begin{figure}[h!tb]
    \centering
    \includegraphics[scale= 0.5, clip= TRUE, trim= 0in 0pt 0in 0in]{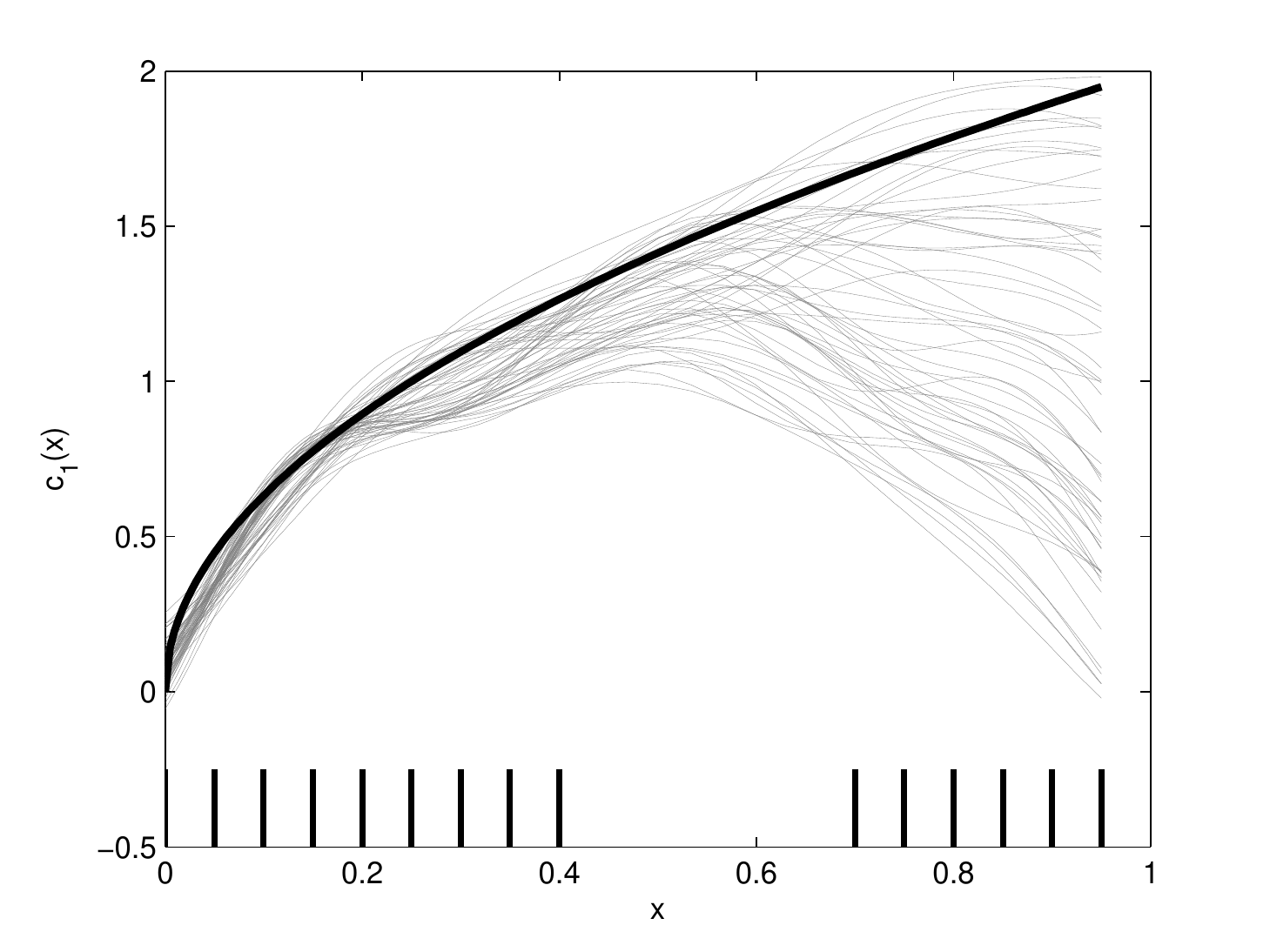}  
    \caption{Sample paths of $c_1(\cdot)$ obtained from combining the three chains in Figure \ref{fig:BadTraces}.}\label{fig:BadPaths}
\end{figure}

Supplementary Figure \ref{fig:fnTauPlots} displays the smoothed approximate posterior densities of $c_2$, $\beta_0^{U}$, and $\beta_1^{U}$. We see that $\beta_0^{U} = 0$ and $\beta_1^{U} = 2$ are contained in the high density regions of their respective posteriors. Thus, in spite of the noninformative prior on $\boldsymbol{\beta}$, we recover the true functional relationship $c_2(x) = 2\sqrt{x}$ with high probability, as evident in the far right panel of the Figure.\\ 
\indent Another situation we consider is the conventional approach in which all calibration parameters are assigned uniform prior distributions over ranges determined from expert opinion. That is, we drop the assumption that $\theta_1$ follows any functional form and suppose it is constant for all $x$. In this case, we take $\pi(c_2) \propto I(2.35 < c_2 < 2.65)$ and $\pi(c_1) \propto I(-0.5 < c_1 < 2.5)$. For the MCMC implementation with the rescaled calibration parameters, we reparameterize the joint posterior in terms of $\log(-\log(\theta_1))$, just as we did with $\theta_2$ to eliminate boundary constraints and facilitate Gaussian proposals for Metropolis sampling. The algorithm then is straightforward.\\
\begin{figure}[tb]
    \centering
    \includegraphics[scale= 0.525, clip= TRUE, trim= 0in 0pt 0in 0in]{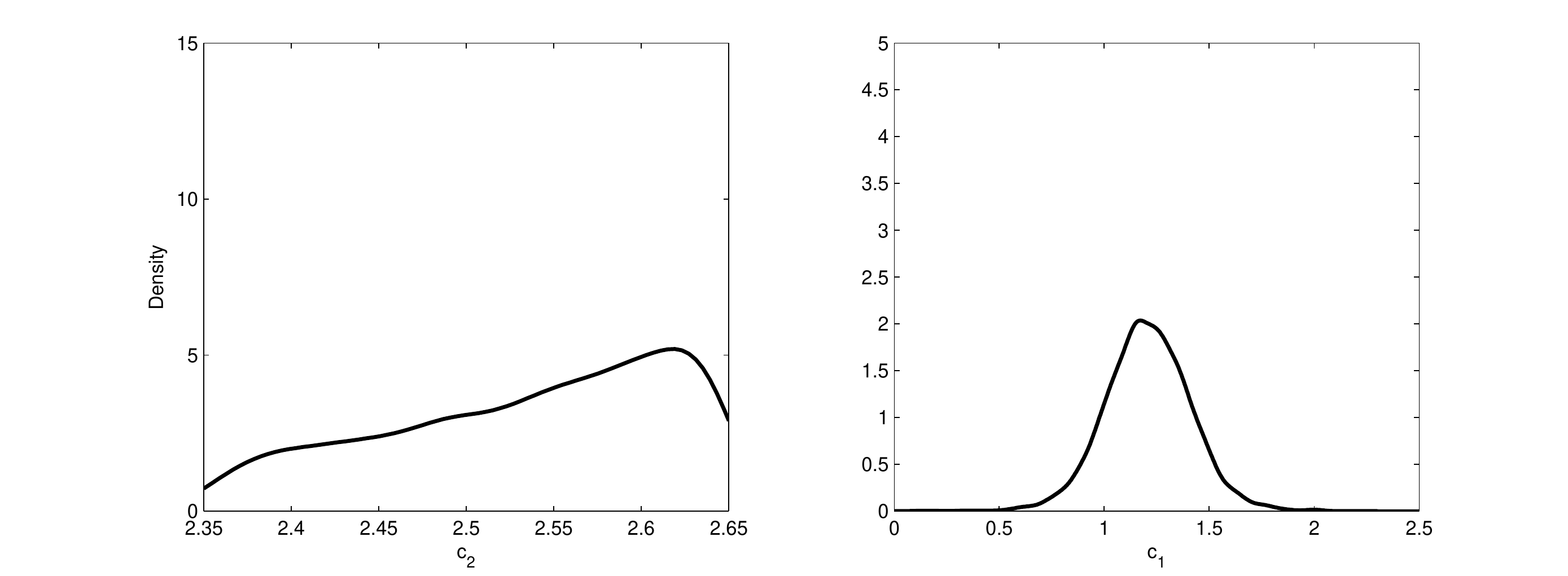}  
    \caption{Smoothed approximate posterior distributions of $c_2$ (left panel) and $c_1$ (right panel) when replacing the GP prior on $\theta_1(\cdot)$ with $\theta_1 \sim \text{Uniform}$ in Model (\ref{eqn:2PModel}).}\label{fig:constTauPlots}
\end{figure}
\indent Figure \ref{fig:constTauPlots} presents the smoothed approximate posterior densities for $c_1$ and $c_2$ resulting from treating both as constant throughout the domain of applicability. Similar to the previous models with informative bounds on $c_2$, we see little additional Bayesian learning about $c_2$. Now, in spite of the overly simplistic treatment of $c_1$, we see considerable posterior concentration about $c_1 \approx 1.25$. This information belies the fact that $c_1$ is truly state-dependent. Thus, we see that strongly identified parameters are no guarantee that the assumed model is the best a researcher can do in describing the system of interest. This could be misleading to the practitioner, who might instead rely on an empirical model discrepancy term to correct the prediction errors seen in Figure \ref{fig:origData}.\\
\indent Supplementary Figure \ref{fig:4CasesPredPlots} plots posterior predictions and approximate 95\% error bars about the holdout design settings for each of the models considered above. While each model is capturing the true responses within its prediction tolerance, an obvious difference between them is in the uncertainties associated with the predictions. As expected, the model assuming the correct functional form for $\theta_1(x)$ results in the best predictions. We see on the other hand that opting for a much more flexible GP model still yields competitive predictions. Little is lost by relaxing the assumption of a specific parametric function. Note the dramatic loss of predictive certainty from treating $\theta_1$ as constant. We quantify the predictive errors with root mean squared predictive error (RMSPE), displayed in Table \ref{tab:rmspe}. The Table shows us that all three models assuming some type of functional dependence vastly outperform the model treating both calibration parameters as constant.\\
\begin{table}[tb]
    \centering
    \begin{tabular}{l | c c c c}
        Model & Parametric $\theta_1(\cdot)$ & Constrained $\theta_1(x_1), ~\theta_1(x_N)$ & Informative $\pi(\theta_2)$ & Constant $\theta_1$ \\
        \hline
        RMSPE & 0.0538 & 0.1185 & 0.0902 & 0.2783\\
    \end{tabular}
    \caption{Root mean squared predictive error (RMSPE) of the posterior predictions at the holdout settings.}\label{tab:rmspe}
\end{table}
\indent Our proposed model allows for a wide variety of link functions. In the GLM framework, the most common link functions for unit interval-valued data are the logit, probit, and cumulative log-log functions. When the values can be assumed to be away from the boundaries with high probability, a Gaussian distributional approximation with the identity link also can be used. Supplementary Figure \ref{fig:4LinksPaths} compares posterior sample paths obtained from our simulated data using each of these link functions with unconstrained sample paths and informative prior bounds on $c_2$. We see that all of them are competitive in terms of recovering the true functional relationship. The differences arise from each link function's effect on the Bayesian learning in the posterior and hence the convergence of the MCMC algorithm. Our experience is that the best convergence is obtained under the identity link. Further, when the Gaussian approximation is justified, faithful posterior estimates of the function are obtained. This approximation also results in the smallest out of sample prediction error, as evident in Table \ref{tab:4LinkRmspe}, which shows the RMSPE for each of the considered link functions. Hence the identity link is preferable, provided the approximation is justified.\\
\begin{table}[tb!]
    \centering
    \begin{tabular}{l | c c c c}
        Link & Logit & Probit & Cumulative Log-Log & Identity \\
        \hline
        RMSPE & 0.0902 & 0.0995 & 0.0957 & 0.0757 \\
    \end{tabular}
    \caption{Root mean squared predictive error (RMSPE) at the holdout settings for each link function.}\label{tab:4LinkRmspe}
\end{table}
\indent To demonstrate what can go wrong, suppose in our example that both $c_2$ and $c_1(\cdot)$ are given very vague priors. The danger here is that weak identifiability might result in highly correlated parameters in the sampling algorithm, since the data are unable to distinguish the effect of one parameter from another. This leads to convergence difficulties in MCMC algorithms. Supplementary Figures \ref{fig:BadTraces} and \ref{fig:BadPaths} display trace plots of the sampled values of $c_2$, $c_1(x_{10})$, and $c_1(x_{15})$ from three different chains using different initial values, along with sample paths of $c_1(\cdot)$ obtained from these chains when using vague priors. The chains do not mix well, indicating convergence problems. Hence, posterior inference is unreliable. When a large sample of field observations is available, the data dominate the posterior so that identifiability is usually not a problem and there is no need to incorporate constraints in the prior model \citep[e.g.,][]{PlumleeEtAl15}. Often in practice, though, the resource-intensive collection of field data limits the available sample size so that using available prior information is crucial to mitigate identifiability problems.\\
\indent Our simulation results demonstrate that our proposed nonparametric functional calibration model can calibrate computer codes and adequately capture unknown functional dependencies between the calibration parameters and the experimental settings.  We see that eliciting such prior information about the parameters can mitigate identifiability problems that are ubiquitous in model validation. Our results suggest the somewhat counterintuitive fact that allowing the calibration parameter $\theta_1$ to vary across the experimental domain results in much less uncertainty about future predictions, in spite of the strong Bayesian learning that occurs when treating $\theta_1$ as constant.  This behavior is particularly appealing since the reduction in uncertainty occurred regardless of whether we imposed the correct functional form or assigned $\theta_1(\cdot)$ a nonparametric Gaussian process prior. We illustrate further that similar results can be obtained regardless of the chosen link function. We emphasize, however, that our experience both in terms of posterior inference as well as convergence of the MCMC algorithm suggests that the identity link approximation is the best option when the true values can be safely assumed to be far from the boundaries with high probability.\\

\section{Application to VPSC Material Plastic Deformation}\label{sec:Application}
As an application, we consider a viscoplastic self-consistent material (VPSC) model for the plastic deformation of polycrystals.  This model, developed by \citet*{LebensohnTome93} and studied by \cite{AtamEtAl14}, treats a polycrystal as a set of single crystals with a texture represented by crystollographic orientations that evolve during plastic deformation. Relationships between deviatoric stress and strain-rate tensors are used to model this viscoplastic deformation. The VPSC formulation imposes a strain-rate during each incremental deformation step, resulting in stress-strain curves as part of the output of the model. The so-called glide-only version of the VPSC model allows dislocations of single crystals to move within the slip plane and hence describes simple shear deformations on this plane. The strain rate at the level of a single crystal, $\dot{\varepsilon}$, is approximated by
\begin{equation}
    \dot{\varepsilon} = \dot{\gamma}_0\sum_{s=1}^{N_s}m^s\left(\frac{|m^s:\sigma|}{\tau_0}\right)^{n_g}\text{sign}(m^s:\sigma),
    \label{eqn:vpscModel}
\end{equation}
where $\sigma$ is the applied stress, $m^s$ is the Schmid tensor, $\tau_0$ is the critical resolved shear stress associated with glide, $n_g$ is the inverse of rate sensitivity for the glide activity, $N_s$ is the total number of active slip systems, $\dot{\gamma}_0$ is a normalizing constant, and $:$ denotes the tensor product.

\citet{StoutEtAl98a, StoutEtAl98b} reported experiments concerning the plastic deformation of 5182 aluminum to which the glide VPSC model is applicable. Two inputs, temperature and strain-rate, were varied in the experiments and stress-strain curves subsequently measured. The experiments were performed until each specimen attained a strain of 0.6, at which time the corresponding stress of the specimen was recorded. Eleven experiments were originally conducted at temperature settings between 200 and 550 $^{\circ}{\rm C}$ and strain-rate equal to $10^{-3}$ and $1$. In the VPSC computer code for implementing (\ref{eqn:vpscModel}), the glide stress exponent $n_g$ and the critical resolved shear stress $\tau_0$ are to be calibrated against the experimental data. Previous empirical work in \citet{AtamEtAl14} suggests that $\tau_0$ is a function of temperature. We thus incorporate this functional dependence into calibrating Model (\ref{eqn:vpscModel}). A parametric functional form was used for $\tau_0(\cdot)$ in \citet{AtamEtAl14}. However, this model is purely empirical in the absence of any existing theory. Hence, we relax the parametric assumption and use a Gaussian process model for $\tau_0(\cdot)$. We use as our field data the experiments conducted at strain-rate equal to $10^{-3}$ while varying temperature. The experimental data are given in Table \ref{tab:fieldData}.
\begin{table}[tb]
    \centering
    \begin{tabular}{l | c c c c c c}
        Experiment & A & B & C & D & E & F\\
        \hline
        Temperature ($^{\circ}{\rm C}$) & 200 & 300 & 350 & 400 & 500 & 550\\
        Maximum Stress (MPa) & 226.2 & 91.4 & 50.0 & 30.6 & 14.9 & 7.0\\
    \end{tabular}
    \caption{Experimental results of plastic deformation of 5812 aluminum \citep*{StoutEtAl98a, StoutEtAl98b}.}\label{tab:fieldData}
\end{table}

\begin{figure}[tb]
    \centering
    \includegraphics[scale= 0.45, clip= TRUE, trim= 0in 0pt 0in 0in]{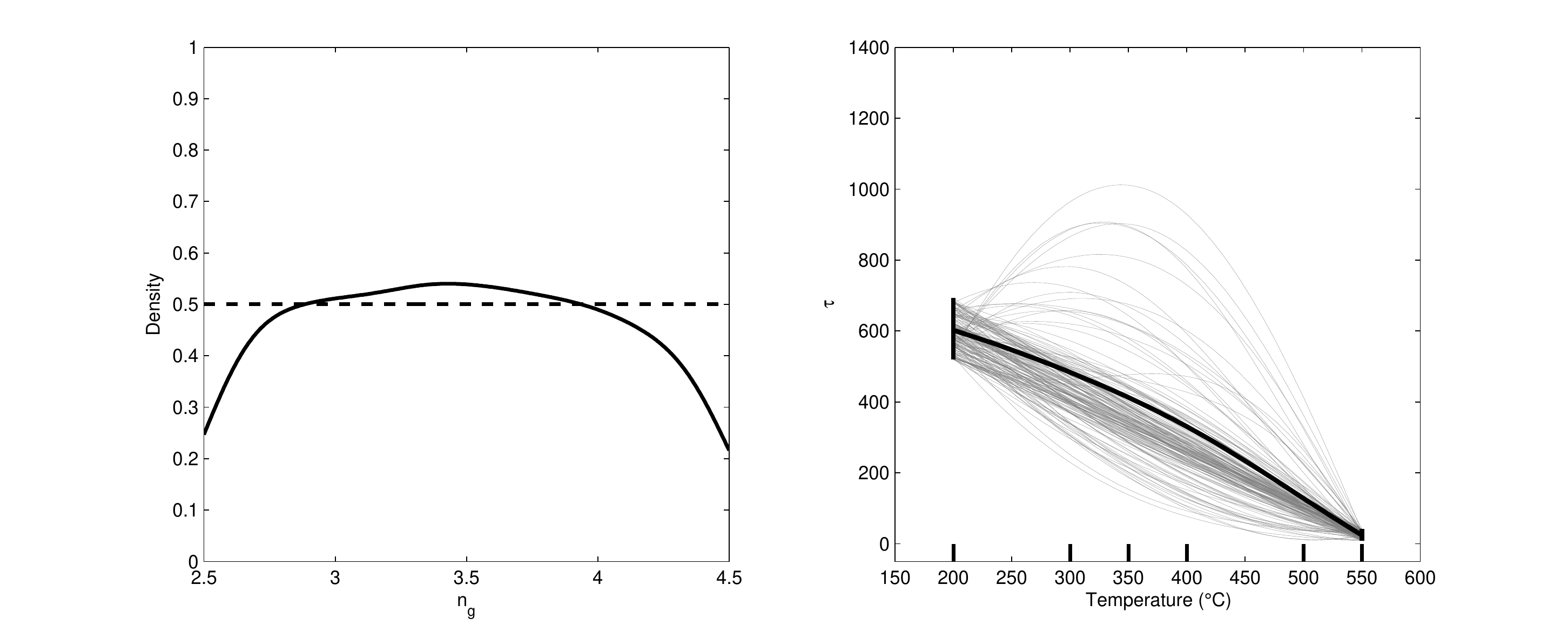}  
    \caption{Smoothed prior and posterior histogram for $n_g$ (left panel) and sample paths drawn from the posterior of $\tau_0(\cdot)$ (right panel). The dark curve in the center is the pointwise mean of the sample paths; the vertical dark lines on the boundaries are the prior constraints imposed on the curves. The larger tick marks along the $x$-axis denote the experimental temperature settings used for the calibration.}\label{fig:VPSCPost}
\end{figure}

\begin{figure}[h!tb]
    \centering
    \includegraphics[scale= 0.55, trim= .5in 0pt 0in 0in]{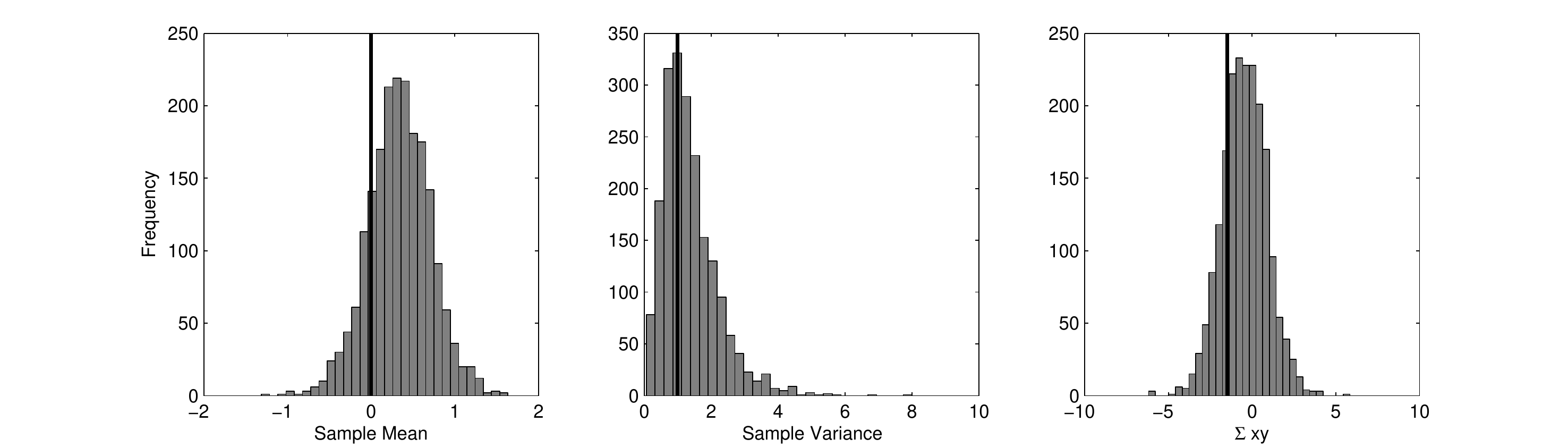}  
    \caption{Histograms of sample statistics calculated from 2,000 replicated datasets from the posterior predictive distribution: $T_1 =$ sample mean (left panel), $T_2 =$ sample variance (middle panel), $T_3 = \sum_{i=1}^N x_iy_i$ (right panel). The dark vertical lines are at the observed statistics from the field data.}\label{fig:VPSCppmc}
\end{figure}

\begin{figure}[h!tb]
    \centering
    \includegraphics[scale= 0.45, clip= TRUE, trim= 0in 0pt 0in 0in]{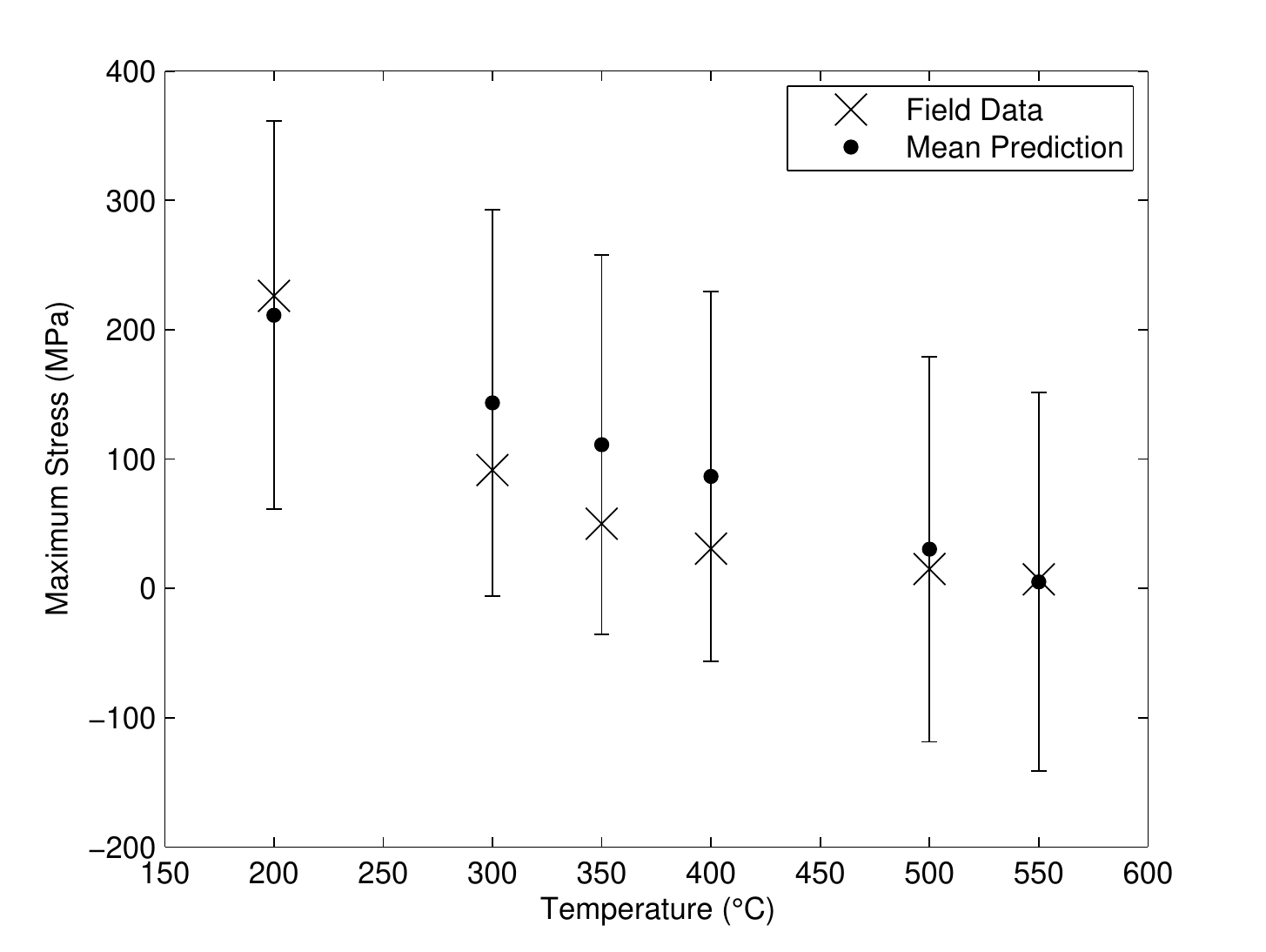}  
    \caption{Posterior predictions of maximum stress from the glide VPSC model with approximate 95\% error bounds at the observed experimental settings.}\label{fig:VPSCPreds}
\end{figure}

We rely on subject matter expert opinion and previous empirical work to determine ranges for possible values of $n_g$ and $\tau_0$. We also have available the extrema for the control input, temperature. These bounds, displayed in Table \ref{tab:bounds}, are used to scale the control and calibration parameters to lie in the unit hypercube prior to calibration. \citet{AtamEtAl14} used nonlinear constrained optimization to obtain optimal values for $\tau_0$ at different temperature settings for use in estimating a parametric function for $\tau_0(\cdot)$. We use this information to refine the constraints on $\tau_0(\cdot)$ at the boundaries of the experimental domain. These values are given in Table \ref{tab:bounds}, as well.\\
\begin{table}[tb]
    \centering
    \begin{tabular}{l | c c c | c c}
        Parameter & Temperature ($^{\circ}{\rm C}$) & $n_g$ & $\tau_0$ (MPa) & $\tau_0(x_1)$ & $\tau_0(x_N)$ \\
        \hline
        Range & [180.00, 570.00] & [2.50, 4.50] & [1.20, 1343.40] & [519.03, 693.07] & [7.78, 42.15]\\
    \end{tabular}
    \caption{Bounds on control and calibration parameters.}\label{tab:bounds}
\end{table}
\indent Figure \ref{fig:VPSCPost} displays the prior and smoothed approximate posterior distribution of $n_g$ along with sample paths drawn from the approximate posterior distribution of $\tau_0(\cdot)$ using the identity link approximation. Superimposed on the sample paths are the pointwise mean curve and the constraints on the boundary values of the paths. For reference, experimental temperature settings used in the calibration are denoted with the large tick marks along the $x$-axis. The density about $n_g$ has updated to become slightly more concentrated about 3.5, in agreement with previous empirical work. The boundary constraints on $\tau_0(\cdot)$ are obviously influential in determining posterior sample paths, as we would expect given the limited experimental data available.\\
\indent As a check of model adequacy, we examine the distributions of selected test quantities of interest, $p(T(\mathbf{y}^{\ast}) \mid \mathbf{y}) = \int p(T(\mathbf{y}^{\ast})\mid \boldsymbol{\theta})\pi(\boldsymbol{\theta} \mid \mathbf{y}) d\boldsymbol{\theta}$, where $\mathbf{y}^{\ast} = (y_1^{\ast}, \ldots, y_N^{\ast})^T$ is a posterior replication of the dataset. The test quantities we use are the sample mean, $T_1(\mathbf{y}) = N^{-1}\sum_{i=1}^N y_i$, the sample variance, $T_2(\mathbf{y}) = (N-1)^{-1}\sum_{i=1}^N(y_i - \overline{y})^2$, and the sample inner product $T_3(\mathbf{y}) = \sum_{i=1}^Nx_iy_i$. These are sufficient statistics for a linear regression of $\mathbf{y}$ on $\mathbf{x}$ and thus summarize salient features of the data. The distributions of these test quantities also enable us to approximate the {\em Bayesian p-values} \citep[][Ch. 6]{GelmanEtAl14}, $p_B^{(i)} = P(T_i(\mathbf{y}^{\ast}) \geq T_i(\mathbf{y}) \mid \mathbf{y}) = \int\int I[T_i(\mathbf{y}^{\ast}) \geq T_i(\mathbf{y})]p(\mathbf{y}^{\ast} \mid \boldsymbol{\theta})\pi(\boldsymbol{\theta} \mid \mathbf{y})d\mathbf{y}^{\ast}d\boldsymbol{\theta}, ~i= 1,2,3$. The Bayesian $p$-value is a simple measure of discrepancy between a hypothesized model and observed data, with values close to zero or one indicating a model's failure to explain features of the data. Supplementary Figure \ref{fig:VPSCppmc} displays histograms of realizations of $T_1$, $T_2$, and $T_3$ based on 2,000 replications drawn from $p(\mathbf{y}^{\ast} \mid \mathbf{y})$. In each plot, the dark vertical line represents the observed value of the statistic from the experimental data. In all three cases, the observed value is well within the range of plausible values posited by our model. The Bayesian $p$-values for each statistic are $p_B^{(1)} = 0.831$, $p_B^{(2)} = 0.616$, and $p_B^{(3)} = 0.785$. Supplementary Figure \ref{fig:VPSCPreds} displays the posterior predictions with approximate 95\% error bars at the observed temperature settings, where we see that the field data are well within the bounds predicted by our model. We can conclude that our modeling assumptions and the subsequent calibrations are entirely consistent with the experimental data.\\
\indent This application illustrates our model's ability to flexibly adapt to changes in appropriate calibration values as a function of the experimental settings while holding other calibration parameters constant. The example also illustrates how the additional uncertainty introduced by omitting the assumption of a parametric functional form is incorporated into model predictions. In the presence of this uncertainty, we still obtain calibrated model predictions that are consistent with field data.\\

\section{Discussion}\label{sec:Conclusions}
Standard practice in computer model calibration is to use expert-elicited prior information to construct relatively simple prior distributions on the calibration parameters and treat them as constant throughout the domain of applicability. While this methodology has proven to be effective, the situation can be improved by acknowledging the fact that the calibrated values might vary as a function of the control inputs and modeling this phenomenon appropriately. Indeed, when models are simplified, the dependence of parameters on the state of the system can be lost. Our proposed nonparametric functional model presented here makes the calibration ``state-aware" through a Gaussian process on the parameters thought to change over the domain.

Through simulation and application, we show that the posterior distribution of our proposed model effectively incorporates prior information and fully accounts for the remaining uncertainty in the presence of small sample sizes while still yielding predictions consistent with experimental observations. We demonstrate that knowing the correct functional form {\em a priori} yields the best predictions with the most precision. However, we are able to obtain competitive predictive performance even after relaxing the parametric function assumption in favor of a nonparametric model.

Our results also suggest that the constant parameter assumption could be misleading in that the posterior distribution may still concentrate around particular calibration parameter values in spite of this assumption being incorrect. In this case, a researcher might opt for a purely empirical model discrepancy term to account for the differences between the calibrated predictions and the field data. Such an approach works well when prediction is the only goal of the calibration procedure. Often, however, inferences about the calibration parameters are desired in addition to reliable prediction of future outcomes. In this case, the presence of a discrepancy term exacerbates identifiability problems that are already present \citep{BayarriEtAl07}. Our proposed approach can obviate the need for an empirical discrepancy term, facilitating stronger inferences while increasing a researcher's confidence in using their model for extrapolation.

Small sample sizes are the norm rather than the exception in computer model calibration, so identifiability is of utmost concern. This paper illustrates that unconstrained functional calibration with vague priors limits posterior inferences. We demonstrate the utility of incorporating prior information which is often available from subject matter experts. Thus, nonparametric functional calibration is still feasible with limited field data. It produces reliable inference and predictions while fully accounting for the uncertainty about the functional form. In addition to boundary constraints, it also might be possible to incorporate prior information such as known monotonicity to further improve identifiability \citep{GolchiEtAl14}.

Our proposed model assumes fast-running computer code, circumventing the need for a surrogate model. It is common in practice, though, for the computer code to be computationally expensive. Indeed, while we are able to obtain the results in Section 4 without an emulator for the VPSC model, the code does in fact take a couple of seconds to execute a single run, making the MCMC routine slow. A natural extension that will be explored in future work is the replacement of the actual computer code in (\ref{eqn:2PModel}) with a surrogate model.  As the dimension of the parameter space increases in a computer model, however, the sensitivities and parameter correlations are much easier to understand when a GP emulator is avoided \citep{HemezAtam11}. We thus recommend using the computer model directly if at all feasible, but acknowledge that the extension of our proposed method to include an emulator is needed.

There remains the question of deciding when to invoke our so-called state-aware calibration, as it may not always be obvious which parameters to treat as functional and which to treat as constant. We suggest beginning with the conventional calibration approach in which all the calibration parameters are treated as constant. The presence of systematic model bias can point to the need for incorporating functional relationships into the calibration. If it is not obvious which parameters might follow a functional relationship, then a sensitivity analysis can be performed, after which the most influential parameters would naturally be the first ones assigned a functional model. Through this approach, a researcher may gain an idea of which parameters to treat as functionally related to the control inputs, but might not know the functional form. At this point, a nonparametric Gaussian process model can be fit to the functional calibration parameters, which then may suggest a specific parametric functional form. Both the parametric and nonparametric versions of the model can be fit and compared using a model assessment tool such as the deviance information criterion \citep[DIC;][]{CarlinLouis09}. If it is found suitable, the parametric model is to be preferred, since it can improve extrapolation and, more importantly, suggest missing physics in the system. State-aware calibration, then, can be a valuable tool for determining when to expand on a currently accepted physics model by revealing previously unknown functional relationships. When found to be consistent with experimentation, suitable parametric functions suggested by the initial nonparametric model will help researchers fill gaps in scientific knowledge.\\

\section{Acknowledgements}
The authors are grateful to Garrison Stevens for her assistance with the \verb|MATLAB| code, to Brian Williams and Cetin Unal of Los Alamos National Laboratory for helpful comments, and to Ricardo Lebensohn of Los Alamos National Laboratory for access to the VPSC code. We thank the editors and referees for their suggestions toward improving this paper, and especially for bringing to our attention the recent work of \citet{PlumleeEtAl15}, which is related to our own work but of which we were previously unaware.

\bibliographystyle{asa}
\bibliography{NPCalibrationRef}

\newpage

\begin{center}
    {\huge Nonparametric Functional Calibration of Computer Models \\ Supplementary Material}\\[18pt]

    {\large D. Andrew Brown \hspace{18pt} Sez Atmaturktur}
\end{center}
Here the reader may find additional material, including the full conditional distributions for implementing the proposed model and supplementary figures.

\section{Full Conditional Distributions for the Proposed Model}
Under the reparamterized version of Model (2.4) we have the following full conditional distributions needed for a Gibbs sampling algorithm:
\begin{equation}
    \begin{aligned}
        \pi(\boldsymbol{\theta}_1^{(\mathbf{x})} \mid \xi, \nu, \lambda_{\theta}, \lambda_y, \mathbf{y}) &\propto \exp\left\{-\frac{\lambda_y}{2}(\mathbf{y} - \boldsymbol{\eta}(\boldsymbol{\theta}_1^{(\mathbf{x})}, \exp\{-e^{\xi}\}))^T(\mathbf{y} - \boldsymbol{\eta}(\boldsymbol{\theta}_1^{(\mathbf{x})}, \exp\{-e^{\xi}\}))\right\} \\
            &~~~~ \times \exp\left\{- \frac{\lambda_{\theta}}{2}(g(\boldsymbol{\theta}_1^{(\mathbf{x})})-\mu_{\theta}\mathbf{1})^T\mathbf{R}_{\nu}^{-1}(g(\boldsymbol{\theta}_1^{(\mathbf{x})})-\mu_{\theta}\mathbf{1})\right\}\\
        \pi(\xi \mid \boldsymbol{\theta}_1^{(\mathbf{x})}, \lambda_y, \mathbf{y}) &\propto \exp\left\{-\frac{\lambda_y}{2}(\mathbf{y} - \boldsymbol{\eta}(\boldsymbol{\theta}_1^{(\mathbf{x})}, \exp\{-e^{\xi}\}))^T(\mathbf{y} - \boldsymbol{\eta}(\boldsymbol{\theta}_1^{(\mathbf{x})}, \exp\{-e^{\xi}\})) + \xi - e^{\xi}\right\}\\
        \lambda_y \mid \boldsymbol{\theta}_1^{(\mathbf{x})}, \xi, \mathbf{y} &\sim \text{Ga}\left(a_y + \frac{N}{2}, ~b_y + \frac{1}{2}(\mathbf{y} - \boldsymbol{\eta}(\boldsymbol{\theta}_1^{(\mathbf{x})}, \exp\{-e^{\xi}\}))^T(\mathbf{y} - \boldsymbol{\eta}(\boldsymbol{\theta}_1^{(\mathbf{x})}, \exp\{-e^{\xi}\}))\right)\\
        \lambda_{\theta} \mid \boldsymbol{\theta}_1^{(\mathbf{x})}, \nu &\sim \text{Ga}\left(a_{\theta} + \frac{N}{2}, ~b_{\theta} + \frac{1}{2}(g(\boldsymbol{\theta}_1^{(\mathbf{x})}) - \mu_{\theta}\mathbf{1})^T\mathbf{R}_{\nu}^{-1}(g(\boldsymbol{\theta}_1^{(\mathbf{x})}) - \mu_{\theta}\mathbf{1})\right)\\
        \pi(\nu \mid \boldsymbol{\theta}_1^{(\mathbf{x})}, \lambda_{\theta}) &\propto |\mathbf{R}_{\nu}|^{-1/2}\exp\left\{-\frac{\lambda_{\theta}}{2}(g(\boldsymbol{\theta}_1^{(\mathbf{x})}) - \mu_{\theta}\mathbf{1})^T\mathbf{R}_{\nu}^{-1}(g(\boldsymbol{\theta}_1^{(\mathbf{x})}) - \mu_{\theta}\mathbf{1}) + \nu - e^{\nu}\right\}\\
            & ~~~~\times (1 - \exp\{-e^{\nu}\})^{b_{\theta} - 1}.
    \end{aligned}
\end{equation}

\newpage

\section{Supplementary Figures}


\end{document}